\documentclass[12pt]{iopart}
\usepackage{amsthm,amssymb,nath}
\usepackage{graphicx}
\def\ax{\xi}
\def\ay{\eta}
\def\gx{X}
\def\gy{Y}

\let\epsilon\varepsilon
\def\hm{\hphantom{-}}
\def\eqref#1{{\rm(\ref{#1})}}
\newtheorem{proposition}{Proposition}

\newtheorem{example}{Example}

\begin{document}

\jl{1}

\title[Integrability of Weingarten surfaces]{On integrability of Weingarten surfaces: a forgotten class%
\footnote{\tiny This is an author-created, un-copyedited version of an article accepted for publication in J. Phys. A: Math. Theor. 42 (2009) 404007. IOP Publishing Ltd is not responsible for any errors or omissions in this version of the manuscript or any version derived from it. The definitive publisher authenticated version is available online at doi:10.1088/1751-8113/42/40/404007.}}
\author{Hynek Baran and Michal Marvan}
\address{Mathematical Institute in Opava, Silesian University in
  Opava, Na Rybn\'\i\v{c}ku 1, 746 01 Opava, Czech Republic.
  {\it E-mail}: Michal.Marvan@math.slu.cz}
\date{}


\begin{abstract}
Rediscovered by a systematic search, a forgotten class of integrable
surfaces is shown to disprove the Finkel--Wu conjecture.
The associated integrable nonlinear partial differential equation
$$
z_{yy} + (1/z)_{xx} + 2 = 0
$$
possesses a zero curvature representation, a third-order symmetry, and a
nonlocal transformation to the sine-Gordon equation
$\phi_{\ax\ay} = \sin\phi$. We leave open the
problem of finding a B\"acklund autotransformation and a recursion 
operator that would produce a local hierarchy.
\end{abstract}

\section{Introduction}

With this paper, we launch a project to classify integrable classes of
surfaces. These are classes of surfaces whose Gauss--Mainardi--Codazzi
equations are integrable in the sense of soliton theory.
Our long-term goals include obtaining lists of integrable classes as
complete as computing resources permit, clarifying their mutual
relations, and identifying known subcases.
Our immediate goal is to demonstrate that the task is feasible and
worth doing.

The classical geometry of immersed surfaces in the Euclidean space is
well known to be closely connected with the modern theory of integrable
systems~\cite{R-S}.
The Gauss--Weingarten equations of a moving frame $\Psi$ always take
the form
$$
\numbered\label{lin problem}
\Psi_x = A \Psi, \quad \Psi_y = B \Psi.
$$
where $A,B$ are appropriate matrix functions. Integrability
conditions of~\eqref{lin problem} are called the
Gauss--Mainardi--Codazzi equations and take the form of a {\it zero
curvature representation}
$$
\numbered\label{ZCR}
A_{y} - B_{x} + [A,B] = 0.
$$
Equation~\eqref{ZCR} is invariant under a huge group of {\it gauge
transformations}
$$
\numbered\label{gauge}
A' = S_x S^{-1} + S A S^{-1}, \qquad B' = S_y S^{-1} + S B S^{-1},
$$
induced by linear transformations $\Psi' = S \Psi$ of the frame.
Here $S$ is an invertible functional matrix, which can be restricted
to take values in the Lie group $G$ associated with the Lie algebra
$\mathfrak g$ matrices $A,B$ belong to -- typically $\mathfrak{so}(3)$.

The zero curvature representation~\eqref{ZCR} is the key ingredient in
the soliton theory~\cite{F-T}, where matrices $A,B$ are additionally
assumed to depend on what is called the {\it spectral parameter}.
The essential requirement for solitonic integrability is that the 
spectral parameter cannot be removed by means of the gauge 
transformation~\eqref{gauge}.
Consequently, if the matrices $A,B$ can be modified so that they depend
on a nonremovable parameter and still satisfy~\eqref{ZCR}, then the
corresponding Gauss--Mainardi--Codazzi equations are considered to be
integrable in the sense of soliton theory, and their solutions are known
as {\it integrable} or {\it soliton surfaces}~\cite{Sym}.

Solitonic integrability can appear only when surfaces are subject to a 
constraint (such as being pseudospherical etc.).
For numerous classical and recent examples see, e.g., the
references~\cite{Bob,R-S,S-K} (or~\cite{Fer} in the projective setting).
Workable tools to classify such constraints include all the general
integrability criteria~\cite{M-S}, which are, however, not immediately
applicable to non-evolutionary systems~\cite{M-N-W}.
Other methods take advantage of the already known non-parametric zero
curvature representation~\eqref{ZCR}, e.g., the method of extended
symmetries by Cie\'sli\'nski et al.~\cite{C-ls,C-nls,C-G-S}.

In this paper we employ a recent method due to one of us~\cite{spp}.
Its essence can be summarized as follows: We attempt to extend the
given non-parametric zero curvature representation (a seed) to a power 
series in terms of the spectral parameter. In the work~\cite{spp}, the
relevant computable cohomological obstructions are identified.
Two obstacles make this procedure not entirely algorithmic: 
The parameter-dependent zero curvature representation can exist only in 
an extension of the Lie algebra $\mathfrak g$ and its jet order (the 
order of derivatives) can exceed that of the seed.
If no obstructions are found, various ways exist to incorporate the 
true nonremovable parameter.

\section{Weingarten surfaces}

To be of genuine interest in geometry, the determining constraint on
integrable surfaces must be invariant with respect to coordinate
changes.
The general non-differential invariant constraint is a functional
relation $f(p,q) = 0$ between the principal curvatures $p,q$.
Such a functional relation is characteristic of Weingarten surfaces,
which have been a topic of continuous interest, especially in global 
differential geometry~\cite{Hop,Vos,S-K,Lop} and computer 
graphics~\cite{vBrunt}.
Well known to be integrable is the class of {\it linear Weingarten
surfaces}~\cite{Dar,R-S}, characterized by a linear relation
$$
\numbered\label{lws}
a k + b h + c = 0, \qquad a,b,c = `const
$$
between the Gauss curvature $k = p q$ and the mean curvature 
$h = \frac 12 (p + q)$ 
(not to be mixed with a linear relation between the principal 
curvatures~\cite{K-S,Lop}).
Other integrable classes of Weingarten surfaces that sporadically
occur in the literature all have a differential defining relation
(e.g., the Hazzidakis equation of the
Bonnet surfaces~\cite{Bob,B-E,Bon}; a harmonicity condition of
Schief's~\cite{Sch} generalized linear Weingarten surfaces) or the
class is not determined by the functional relation $f(p,q) = 0$
alone (e.g.,~\cite{C-F-G}).

So far, nothing contradicts the conjecture of 
Finkel~\cite[Conjecture~3.4]{Fin} and Wu~\cite{Wu} that the only 
functional relation $f(p,q) = 0$ to determine an integrable class 
of Weingarten surfaces is the linear relation~\eqref{lws}.
Supporting arguments include Wu's~\cite{Wu} proof of non-existence 
of an $\mathfrak{so}(3)$-valued zero-curvature representation 
depending only on $x$-derivatives.
Finkel's~\cite{Fin} argument roots in an unsuccessful search for
higher-order symmetries and a (disputable, see~\cite[\S2]{M-N-W}) 
conjecture that integrability implies the existence of a local 
higher-order symmetry (actually the infinite hierarchy can be 
nonlocal, see also~\cite[\S1.4.4.2]{M-S}).

Nevertheless, the main result of the present paper asserts that the
simple relation
$$
\numbered\label{nws1}
\frac 1p - \frac 1q = `const
$$
between the main curvatures $p,q$, determines an integrable class 
of Weingarten surfaces.
The associated nonlinear partial differential equation~\eqref{z} has
a parameter-dependent zero curvature representation~\eqref{z:zcr}
(outside the class considered in~\cite{Wu}), 
a third-order symmetry~\eqref{z:sym} (missed in~\cite{Fin}), 
and a recursion operator~\eqref{z:ro}.

Paradoxically enough, surfaces satisfying relation~\eqref{nws1} were
not unknown to nine\-teenth century geometers.
In view of their knowledge, our integrability result is not an 
entirely unexpected one.
In fact, Ribaucour~\cite{Ri} established that the corresponding 
focal surfaces (evolutes) have a constant Gaussian curvature $k < 0$
(are pseudospherical).
Conversely, surfaces satisfying equation~\eqref{nws1} are involutes
of pseudospherical surfaces. Moreover, the classical Bianchi
transformation~\cite{Bia} is nothing but the induced correspondence
between the two focal pseudospherical surfaces.
Ribaucour's theorems are covered in Darboux~\cite{Dar} and early
twentieth-century monographs, such as~\cite{BiaI,Eis,For,Wea}.
Later they became obsolete and forgotten as the induced Bianchi relation
between pseudo\-spherical surfaces became superseded by the classical
B\"acklund transformation (the history is nicely reviewed by Prus and
Sym in~\cite[Sect.~4]{P-S}).

The first examples of surfaces satisfying relation~\eqref{nws1} also
date to the nineteenth century.
Lipschitz~\cite{Lip} derived a four-parametric family in terms of
elliptic integrals.
A particular subcase, the rotation surface of von Lilienthal~\cite{Lil},
is the involute surface of the pseudosphere.

The left-hand side of Equation~\eqref{nws1} is equal to the difference
of the principal radii of curvature at a point.
This geometric quantity has a definite physical meaning, being
associated with the {\it interval of Sturm}~\cite{Sturm}, also
known as the {\it astigmatic interval\/} or the
{\it amplitude of astigmatism\/} or simply the
{\it astigmatism}~\cite{Gr-I}.
A mirror or a refracting surface satisfying relation~\eqref{nws1} will
feature a constant amplitude of astigmatism in the normal directions.
In the sequel, surfaces satisfying condition~\eqref{nws1} will be called 
{\it surfaces of constant astigmatism.}
Accordingly, the equation~\eqref{z} to determine the surfaces of constant 
astigmatism will be called the {\it constant astigmatism equation}.

\section{Preliminaries}
\label{sect:prelim}

We shall consider surfaces $\mathbf r(x,y)$ parametrized by curvature
lines.
As is well known, the fundamental forms can be written as
$$
`I = u^2\,`d x^2 + v^2\,`d y^2,  \\
`II = u^2 p\,`d x^2 + v^2 q\,`d y^2,
$$
where $p,q$ are the principal curvatures.
Coordinates $x,y$ are unique up to arbitrary changes $x = X(x)$,
$y = Y(y)$.

Let $\Psi = (\mathbf e_1,\mathbf e_2,\mathbf n)$ denote the orthonormal
frame, given by
$\mathbf e_1 = \mathbf r_x/u$,
$\mathbf e_2 = \mathbf r_y/v$,
$\mathbf n = \mathbf e_1 \times \mathbf e_2$.
The Gauss--Weingarten equations
$$
\numbered \label{G-W:so3}
\Psi_x = \left(\begin{array}{ccc}
     \hm 0 & -\frac{u_y}{v} & \hm u p \\
     \hm \frac{u_y}{v} & \hm 0 & \hm 0 \\
     -u p & \hm 0 & \hm 0 \end{array}\right) \Psi,
\quad
\Psi_y = \left(\begin{array}{ccc}
     \hm 0 & \hm \frac{v_x}{u} & \hm 0 \\
     -\frac{v_x}{u} & \hm 0 & \hm v q \\
     \hm 0 & -v q & \hm 0 \end{array}\right) \Psi.
$$
are easily established. Their integrability conditions are the Gauss 
equation
\begin{equation} \label{GMC-G}
u u_{yy} + v v_{xx} - \frac{v}{u} u_x v_x - \frac{u}{v} u_y v_y
     + u^2 v^2 p q = 0,
\end{equation}
and the Mainardi--Codazzi equations
\begin{equation}
(p - q) u_y + u p_y = 0, \quad
(q - p) v_x + v q_x = 0. \label{GMC-MC}
\end{equation}
Consequently, the two~$\mathfrak{so}(3)$ matrices occurring in
formulas~\eqref{G-W:so3} constitute a nonparametric zero curvature
representation of the Gauss--Mainardi--Codazzi 
system~\eqref{GMC-G},~\eqref{GMC-MC}.
Because of the isomorphism $\mathfrak{so}(3,\mathbb C) \cong
\mathfrak{sl}(2,\mathbb C)$, the same zero curvature representation 
can be alternatively written in terms of $2 \times 2$ matrices
$$
\numbered \label{0}
A_0 = \left(\begin{array}{cc}
     \frac{`i u_y}{2v} &
     -\frac{1}{2} u p \\
     \frac{1}{2} u p &
     -\frac{`i u_y}{2v} \end{array}\right),
\qquad
B_0 = \left(\begin{array}{cc}
     -\frac{`i v_x}{2u} &
     -\frac{1}{2} `i q v \\
     -\frac{1}{2} `i q v &
     \hm \frac{`i v_x}{2u}
\end{array}\right).
$$

Let us impose a constraint~$f(p,q) = 0$.
If nontrivial, it can be resolved with respect to one of
the curvatures, say
$$
\numbered\label{q=F(p)}
q = F(p),
$$
which we assume henceforth.
Then the Gauss--Mainardi--Codazzi system reduces
substantially~\cite{vBrunt,Fin,Wu}.
In particular, the Mainardi--Codazzi equations~\eqref{GMC-MC} have a
general solution
$$
u = \frac{u_0}E, \quad
v = -v_0 E', \quad
q = p - \frac{E}{E'},
$$
where $E = E(p)$ is an arbitrary nonconstant function, $E' = dE/dp$, and
and $u_0,v_0$ are 
functions of $x$ and $y$, respectively, removable by transformation 
$\tilde x = \int u_0\,dx$, $\tilde y = \int v_0\,dy$.
Therefore, we can put $u_0 = -v_0 = 1$ without loss of generality, i.e.,
$$
\numbered\label{E2uvq}
u = \frac1E, \quad
v = E', \quad
q = p - \frac{E}{E'}.
$$
The Gauss equation~\eqref{GMC-G} then becomes
\begin{equation}
\label{G}
p_{yy} =
E^3 E'' p_{xx}
     + 2 \frac{E'}E p_y^2
     + E^2 (E E'')' p_x^2
     + E E' p^2 - E^2 p.
\end{equation}
Summarizing, the Gauss--Mainardi--Codazzi system of Weingarten
surfaces reduces to the single equation~\eqref{G}.
The classification problem considered in this paper is 
``for which choices of the function $E(p)$ is the equation~\eqref{G} 
integrable?''

By substituting~\eqref{E2uvq} into \eqref{0}, we easily obtain a
nonparametric zero curvature representation of equation~\eqref{G},
$$
\numbered\label{AB0}
A_0 = (\begin{array}{cc}
\frac{`i}2 \frac{p_y}{E^2} & -\frac12 \frac p E \\
\frac12 \frac p E & -\frac{`i}2 \frac{p_y}{E^2}
\end{array}), \quad
B_0 = (\begin{array}{cc}
\frac{`i}2 E E'' p_x & \frac{`i}2 (E' p - E) \\
\frac{`i}2 (E' p - E) & -\frac{`i}2 E E'' p_x
\end{array}),
$$
which will be the starting point of the calculations to follow.

\section{Cohomological criteria}

Readers not interested in details of the classification method can
skip this section and continue to investigation of surfaces of
constant astigmatism in Section~\ref{z:sect}.

We use the formal theory of partial differential equations, which
treats coordinates, unknown functions, and their derivatives as
independent quantities.
Equations can be conveniently represented as submanifolds in
appropriate jet spaces~\cite{B-V-V}. All our considerations being
local, we let $J^\infty = J^\infty(\mathbb R^2, \mathbb R)$ denote the
space of $\infty$-jets of smooth functions $\mathbb R^2 \to \mathbb R$.
The base $\mathbb R^2$ being equipped with coordinates $x,y$, the
natural coordinates along fibres of $J^\infty \to \mathbb R^2$
correspond to $p$ and its derivatives. These will be denoted $p_I$,
where $I$ stands for a symmetric multiindex in $x,y$ 
(including the ``empty'' multiindex $\emptyset$ such that 
$p_\emptyset = p$).
The usual total derivatives
$$
D_x = \frac\partial{\partial x}
     + \sum_{I} p_{xI} \frac{\partial}{\partial p_I},
\qquad
D_y = \frac\partial{\partial y}
     + \sum_{I} p_{yI} \frac{\partial}{\partial p_I}
$$
can be viewed as acting on smooth functions defined on $J^\infty$
(by definition, a smooth function locally depends on a finite number
of coordinates).

In $J^\infty$, we consider a submanifold $\mathcal G$ determined by
equation~\eqref{G} and all its differential consequences obtained by
taking successive total derivatives of both sides of~\eqref{G}.
On $\mathcal G$, all derivatives of the form $p_{Jyy}$ become
expressible in terms of the others.
Therefore, derivatives $p_I$ with $y$ occurring no more than twice in
$I$ serve as natural coordinates along the fibres of $\mathcal G \to
\mathbb R^2$.
Being tangent to~\eqref{G}, the total derivatives admit a restriction
to $\mathcal G$. We retain the same notation $D_x,D_y$ for the
restricted total derivatives.

The essence of the adopted point of view can be summarized as follows:
A function $f$ on $J^\infty$ satisfies $f|_{\mathcal G} = 0$ if and
only if $f$ is zero as a consequence of equation~\eqref{G}.
  From now on we assume that all objects (like the matrices $A,B$) are
defined on $\mathcal G$.
When writing
$$
\numbered\label{ZCR'}
\left.(D_{y} A - D_{x} B + [A,B])\right|_{\mathcal G} = 0
$$
we mean that the zero curvature condition~\eqref{ZCR} holds as a
consequence of equation~\eqref{G}.

In what follows, characteristic elements~\cite{M1,M2,Sak} play a crucial
role. These are nonabelian analogues of characteristics of conservation
laws~\cite{B-V-V}.
For instance, the characteristic element of the initial zero curvature
representation~\eqref{AB0} is the $\mathfrak{sl}(2,\mathbb C)$-matrix
$$
C_0 = (\begin{array}{cc}
\frac{`i}2 \frac 1{E^2} & 0 \\ 0 & -\frac{`i}2 \frac 1{E^2}
\end{array}).
$$
This immediately follows from the fact that
$$  
D_{y} A_0 - D_{x} B_0 + [A_0,B_0] = C_0 F,
$$
where
$$
F = p_{yy} -
E^3 E'' p_{xx}
     - 2 \frac{E'}E p_y^2
     - E^2 (E E'')' p_x^2
     - p^2 E E' - p E^2,
$$
so that the Gauss equation~\eqref{G} can be written as $F = 0$.

Let $A = A(\lambda)$, $B = B(\lambda)$ be the parametric zero curvature
representation sought, $C = C(\lambda)$ the corresponding
characteristic element. Besides~\eqref{ZCR'}, they will also satisfy the
formula~\cite{M1}
$$
\numbered\label{ds}
\left.
   \sum_{I} (-\hat D)_I (\frac{\partial F}{\partial u^k_I} C)
\right|_{\mathcal G} = 0,
$$
with $I$ running over all symmetric multiindices, including the empty
one. Here $\hat D_x = D_x - [A,\adot]$, $\hat D_y = D_y - [B,\adot]$,
the other values being obtained by composition, which can be taken in
any order since~\eqref{ZCR'} implies that $\hat D_x, \hat D_y$ commute.

Characteristic elements of gauge equivalent zero curvature
representations are conjugate (similar). This allows us to transform
characteristic elements into the normal form with respect to
conjugation,
namely, the Jordan normal form.
Since the matrix $C_0$ above is diagonal, it follows that for $\lambda$
sufficiently close to zero the characteristic element $C(\lambda)$ will
be also diagonalizable.

However, diagonal matrices have a nontrivial stabilizer
$\mathcal S \subset `SL(2,\mathbb C)$
with respect to conjugation, which consists of diagonal matrices
$$
(\begin{array}{cc} s & 0 \\ 0 & 1/s \end{array}).
$$
Gauge transformations from the group $\mathcal S$
(henceforth $\mathcal S$-transformations)
preserve the characteristic elements $C(\lambda)$.
Their gauge action on a general $\mathfrak{sl}(2)$-valued zero
curvature representation $A,B$ is sufficiently simple:
$$
(\begin{array}{cc} a_{11} & a_{12} \\ a_{21} & -a_{11} \end{array})
\mapsto
(\begin{array}{cc} \frac{s_x}s + a_{11} & s^2 a_{12} \\
    \frac{a_{21}}{s^2} & -\frac{s_x}s - a_{11} \end{array})
$$
and similarly for $B$. Using $\mathcal S$-transformations, one can
achieve a unique normal form of matrices $A,B$ as follows:
If $a_{12} \ne 0$, then by setting $s = (a_{21}/a_{12})^{1/4}$
we turn $A$ into a symmetric matrix, while in the remaining case
$a_{12} = 0$ the zero curvature representation degenerates to a pair
of conservation laws~\cite{M2}.
In other words, being symmetric is a normal form of nondegenerate zero
curvature representations with respect to $\mathcal S$-transformations.

Turning back to our original problem, we see that $B_0$ is symmetric,
and therefore the nearby matrices $B(\lambda)$ can also be symmetrized
by an $\mathcal S$-transformation.
A simple calculation shows that, by assuming diagonality of $C(\lambda)$
and symmetricity of $B(\lambda)$, we make the system~\eqref{ds}
determined, hence solvable (actually, we fix the gauge).

Summarizing, the computation of zero curvature representation has been
reduced to solution of the determined system~\eqref{ZCR'}, \eqref{ds}
under a suitable choice of normal forms for $C$ and $B$. However,
this nonlinear system is still quite difficult to solve even with the
help of computer algebra. 
To linearize the system, the work~\cite{spp} considers Taylor expansions
$$
\numbered\label{Tayl}
A(\lambda) = \sum_{k = 0} A_k \lambda^k, \quad
B(\lambda) = \sum_{k = 0} B_k \lambda^k, \quad
C(\lambda) = \sum_{k = 0} C_k \lambda^k,
$$
with $A_0,B_0,C_0$ coming from the initial parameterless zero curvature
representation~\eqref{0}.
The condition of zero curvature for $A(\lambda),B(\lambda)$ implies an
infinite sequence of conditions of zero curvature for block triangular 
matrices
$$
\numbered\label{barbar AB}
A^{[m]} = (\begin{array}{cccc} A_0 & 0 & \dots & 0 \\
A_1 & A_0 & \ddots & \vdots \\
\vdots & \ddots & \ddots & 0 \\
A_m & \dots & A_1 & A_0 \end{array}),
\quad
B^{[m]} = (\begin{array}{cccc} B_0 & 0 & \dots & 0 \\
B_1 & B_0 & \ddots & \vdots \\
\vdots & \ddots & \ddots & 0 \\
B_m & \dots & B_1 & B_0 \end{array}).
$$

Characteristic elements $C^{[m]}$ assume the same form.
Zero curvature representations $A^{[m]},B^{[m]}$ are to be considered
under the gauge group consisting of block triangular matrices
$$
S^{[m]} = (\begin{array}{cccc}
E & 0 & \dots & 0 \\
S_1 & E & \ddots & \vdots \\
\vdots & \ddots & \ddots & 0 \\
S_m & \dots & S_1 & E \end{array}).
$$
with unit matrices $E$ in the diagonal positions.
By a cohomological argument presented in~\cite[Prop.~1]{spp}, a 
nontrivial family $A(\lambda),B(\lambda)$ with analytic dependence
on $\lambda$ has expansions~\eqref{Tayl} such that $A_1$ or $B_1$ is 
not zero.

Let \eqref{ZCR'}$^{[m]}$,~\eqref{ds}$^{[m]}$ denote the
system obtained by substituting $A \to A^{[m]}, B \to B^{[m]}$ into
system~\eqref{ZCR'},~\eqref{ds}, for arbitrary $m > 0$.
Observe that systems \eqref{ZCR'}$^{[m]}$,~\eqref{ds}$^{[m]}$
are linear in 
their highest order unknowns $A_m,B_m,C_m$ and can be solved 
sequentially.
Then the applicable cohomological criterion can be
summarized as follows.

\begin{proposition}[{\cite[Prop.~3]{spp}}]
\label{cohom}
Let $m > 0$. 
If $A_1 = B_1 = 0$ for all solutions $A^{[m]},B^{[m]}$ of 
system~\eqref{ZCR'}$^{[m]}$,~\eqref{ds}$^{[m]}$, then there is no 
possibility to construct expansions~\eqref{Tayl} of order~$m$, and 
consequently, the seed zero curvature representation $A_0,B_0$ cannot 
belong to a nontrivial analytic family.
\end{proposition}

Finally, to be able to solve
system~\eqref{ZCR'}$^{[m]}$,~\eqref{ds}$^{[m]}$,
we need to know the normal forms of matrices $A^{[m]},B^{[m]}$.
However, the normal forms for $B(\lambda),C(\lambda)$ established above
immediately imply the same normal forms for $C_k$ (diagonal) and
$B_k$ (symmetric).

\section{Results}
\label{z:sect}

In this section, we present the results of computation of the
cohomological obstructions in the case of the nonparametric zero
curvature representation~\eqref{AB0} of equation~\eqref{G}.
As a sub-result we obtain the first few coefficients $A_k,B_k$ of
Taylor expansions~\eqref{Tayl}.

As we have seen in the preceding section (Proposition~\ref{cohom}),
the problem reduces to solving the
system~\eqref{ZCR'}$^{[m]}$,~\eqref{ds}$^{[m]}$ of linear differential
equations in total derivatives, for increasing values of~$m$.
This is only possible under a suitable restriction on the jet order
of the unknowns $A_k,B_k,C_k$, $k > 0$. To start with, we assume
dependence on the first-order jets at most.
Upon expanding all total derivatives,
equations~\eqref{ZCR'}$^{[m]}$,~\eqref{ds}$^{[m]}$ become a large 
overdetermined system of linear partial differential equations.
As such, the system is solvable by computing the passive (or involutive)
form under a suitable (elimination) ranking~\cite{R-L-W}.

Starting with $m = 1$, we checked that nonzero matrices $A_1,B_1$ 
depending on second order derivatives exist for all possible determining
relations~\eqref{q=F(p)}. When incrementing $m$ to $2$, nontrivial 
conditions started to appear, but we also reached the boundaries of our 
available computing resources. Consequently, our present classification
results are still incomplete.
Nevertheless, we were able to obtain a passive system of differential
equations in several cases. Moreover, in two cases we were able to
find $A_2,B_2$ explicitly. One of them were the linear
Weingarten surfaces~\eqref{lws}.
Their integrability is a well established fact~\cite{R-S}, the
associated sine-Gordon equation $\phi_{xy} = \sin\phi$ being a
textbook example of integrability.
The other class emerged as a solution
\begin{equation}
\label{E}
E = \frac{p}{`e^{1 + c/p}}, \qquad c = `const,
\end{equation}
of the ordinary differential equation
$$
\frac{E''}E
     - (\frac{E'} E)^2
     + \frac 2 p \frac{E'}{E}
     - \frac 1{p^2}
= 0.
$$

Henceforth we concentrate on the solution~\eqref{E}.
The coefficients $u,v,q$ are easily found from~\eqref{E2uvq}
to be
$$
u = \frac{`e^{1 + c/p}}{p}, \quad
v = \frac{p + c}{p `e^{1 + c/p}}, \quad
q = \frac{p c}{p + c}.
$$
The last equality shows that the condition of constant
astigmatism~\eqref{nws1} holds with the constant $-1/c$ on the 
right-hand side. The Gauss equation~\eqref{G} becomes
$$
p_{yy} = \frac{c^2}{`e^{\displayed{4 (1 + \frac cp)}}} p_{xx}
     + 2 \frac{p + c}{p^2} p_y^2
     + 2 \frac{c^2(c - p)}{`e^{\displayed{4 (1 + \frac cp)}} p^2} p_x^2
     + \frac{c p^2}{`e^{\displayed{2 (1 + \frac cp)}}}.
$$

In principle, the cohomological method we applied can only prove
nonintegrability and only indicate, but not prove, integrability.
However, it was easy to guess an ansatz based on the
form of $A_k$ and $B_k$. By solving~\eqref{ZCR'},~\eqref{ds}
we obtained a $\lambda$-dependent zero curvature representation
\begin{equation}
\label{p:zcr}
A =
(\begin{array}{cc}
\lambda c \frac{p_x}{p^2} + \sqrt{\lambda^2 + \lambda}
       `e^{\displayed{2 (1 + \frac{c}{p})}} \frac{p_y}{p^2} &
\lambda `e^{\displayed{1 + 2 \frac{c}{p}}} \\
(\lambda + 1) `e &
-\lambda c \frac{p_x}{p^2} - \sqrt{\lambda^2 + \lambda}
       `e^{\displayed{2 (1 + \frac{c}{p})}} \frac{p_y}{p^2}
\end{array}),
\\
B =
(\begin{array}{cc}
\lambda c \frac{p_y}{p^2} + \sqrt{\lambda^2 + \lambda} c^2 
       `e^{\displayed{-2 (1 + \frac{c}{p})}} \frac{p_x}{p^2} &
\sqrt{\lambda^2 + \lambda} c `e^{-1}
\\
\sqrt{\lambda^2 + \lambda} c 
       `e^{\displayed{-1 - 2\frac{c}{p}}} &
-\lambda c \frac{p_y}{p^2} - \sqrt{\lambda^2 + \lambda} c^2
       `e^{-\displayed{2 (1 + \frac{c}{p})}} \frac{p_x}{p^2}
\end{array}),
\end{equation}
which reduces to the initial $A_0,B_0$ given by~\eqref{AB0}
when $\lambda = -\frac12$.
The dependence on $p_y$ explains why this class of 
Weingarten surfaces is missing in Wu's paper~\cite{Wu}.

Upon substitution
\begin{equation}
\label{p2z}
x \to \frac x{|c|^{1/4}}, \quad
y \to \frac y{|c|^{3/4}}, \quad
p \to \frac {4 c}{2 \ln z + \ln |c| - 4}
\end{equation}
the Gauss equation~\eqref{G} simplifies to
$$
\numbered\label{z}
z_{yy} + (\frac1z)_{xx} + 2 = 0,
$$
and the zero-curvature representation~\eqref{p:zcr} to
\begin{equation}
\label{z:zcr}
A =
(\begin{array}{cc}
\frac12 \sqrt{\lambda^2 + \lambda} z_y + \frac{1 + 2\lambda}{4} \frac{z_x}{z} &
(\lambda + 1) \sqrt z \\
\lambda \sqrt z &
-\frac12 \sqrt{\lambda^2 + \lambda} z_y - \frac{1 + 2\lambda}{4} \frac{z_x}{z}
\end{array}), \\
B =
(\begin{array}{cc}
\frac12 \sqrt{\lambda^2 + \lambda} \frac{z_x}{z^2}
 + \frac{1 + 2\lambda}{4} \frac{z_y}{z} &
\frac{\sqrt{\lambda^2 + \lambda}}{\sqrt z} \\
\frac{\sqrt{\lambda^2 + \lambda}}{\sqrt z} &
-\frac12 \sqrt{\lambda^2 + \lambda} \frac{z_x}{z^2} 
 - \frac{1 + 2\lambda}{4} \frac{z_y}{z}
\end{array}).
\end{equation}
Let us remark that one can remove the $x$-derivatives from $A$ and $y$-derivatives 
from $B$ by the gauge transformation~\eqref{gauge}, albeit at the cost of introducing an exponential dependence on the spectral parameter. 
In~\eqref{p:zcr} and~\eqref{z:zcr}, the corresponding gauge matrix is
$$
S = (\begin{array}{cc}
`e^{-\frac{\lambda c}{p}} & 0
\\
0 &
`e^{\frac{\lambda c}{p}}
\end{array})
\quad\text{and}\quad
S = (\begin{array}{cc}
z^{\lambda/2} & 0
\\
0 &
z^{-\lambda/2}
\end{array}),
$$
respectively.
For instance, the pair~\eqref{z:zcr} becomes
$$
A' =
(\begin{array}{cc}
\frac12 \sqrt{\lambda^2 + \lambda} z_y &
(\lambda + 1) z^{-\lambda} \\
\lambda z^{\lambda + 1} &
-\frac12 \sqrt{\lambda^2 + \lambda} z_y
\end{array}), \quad
B' =
(\begin{array}{cc}
\left.\frac12\right. \sqrt{\lambda^2 + \lambda} \frac{z_x}{z^2} &
\sqrt{\lambda^2 + \lambda} z^{-\lambda - 1} \\
\sqrt{\lambda^2 + \lambda} z^\lambda &
-\left.\frac12\right. \sqrt{\lambda^2 + \lambda} \frac{z_x}{z^2}
\end{array}).
$$

Equation~\eqref{z} has obvious translational symmetries
$\partial_x, \partial_y$,
the scaling symmetry $2 z \partial_z - x \partial_x + y \partial_y$,
and a discrete symmetry
$$
\numbered\label{z:ds}
x \to y, \quad y \to x, \quad z \to \frac1z.
$$
Computation reveals also two third-order symmetries of 
equation~\eqref{z}.
One of them has the generator
$$
\numbered\label{z:sym}
\padded{\quad}
\frac{z^3}{K^3} (z_{xxx} - z z_{xxy})
   - \frac3{K^5} z^3 (z_x - z z_y) (z_{xx} - z z_{xy})^2
\\  - \frac2{K^5} z^5 (9 z_x - z z_y) z_{xx}
    + \frac1{2 K^5} z^2 (9 z_x^2 + 4 z z_x z_y - z^2 z_y^2) (z_x - z z_y)
z_{xx}
\\  - \frac2{K^5} z^3 z_x (z_x - z z_y) (4 z_x - z z_y) z_{xy}
    + \frac4{K^5} z^6 z_x z_{xy}
\\  + \frac3{K^5} z^4 (5 z_x - z z_y) z_x^2
    - \frac3{K^5} z (z_x - z z_y) z_x^4,
\return
$$
where
$$
K = \sqrt{(z_x - z z_y)^2 + 4 z^3}.
$$
The other is obtained by conjugation with the discrete
symmetry~\eqref{z:ds}.

Moreover, A.~Sergyeyev succeeded in finding a recursion operator
for equation~\eqref{z} in the usual pseudodifferential form
$$
\numbered\label{z:ro}
-z_y D_x^{-1} + z_x D_x^{-2} D_y + 2 z D_x^{-1} D_y
$$
(unpublished).
As far as we could see, the operator generates only nonlocal symmetries.
We leave as an open problem to find a recursion operator that would
generate the third-order symmetry~\eqref{z:sym}.

Let us conclude this section with some easy geometric observations.
First of all, we can put $c = 1$ without loss of generality. 
This can be always achieved by rescaling the ambient Euclidean metric 
and, if necessary, changing the orientation. 

Now, the symmetries of the constant astigmatism equation~\eqref{z} 
have the following geometric interpretation.
Translation symmetries are simply reparametrizations of the surface.
The scaling symmetry $\phi_\epsilon$: $x \to `e^\epsilon x$, 
$y \to `e^{-\epsilon} y$, $z \to `e^{-2\epsilon} z$ takes a given 
surface $\mathbf r(x,y)$ to the parallel surface 
$\mathbf r(x,y) + \epsilon \mathbf n(x,y)$.
This is not surprising since parallel surfaces obviously have equal 
astigmatism in the corresponding points.
Finally, swapping the orientation is another symmetry, 
which can be identified with a composition of the discrete 
symmetry~\eqref{z:ds} and the rescaling $\phi_1$.
Hence, the discrete symmetry~\eqref{z:ds} corresponds to the 
change of the orientation followed by taking the parallel 
surface at the unit distance.

\section{Relation to pseudospherical surfaces}

As already mentioned in the introduction, nineteenth century geometers
knew of a simple relation between pseudospheric surfaces and surfaces
of constant astigmatism, even though they did not find the latter
important enough to be named.
In this section we reproduce some of their findings and derive a nonlocal 
transformation between the constant astigmatism equation~\eqref{z} and 
the famous sine-Gordon equation.
Again, we put $c = 1$ for simplicity, meaning that the associated focal 
surfaces will be of Gaussian curvature~$-1$.

The forthcoming calculations are conveniently performed in terms of
the variable $z$ given by formula~\eqref{p2z} or a new variable $w$ 
related to $z$ by
$$
\numbered\label{z2w}
z = `e^{2 w}.
$$
Then we have
$$
\numbered\label{pquv2w}
u = (w - 1) `e^w, \quad
v = \frac{w}{`e^w}, \quad
p = \frac{1}{w - 1}, \quad
q = \frac{1}{w}.
$$
and the discrete symmetry~\eqref{z:ds} becomes simply
$$
\numbered\label{w:ds}
x \to y, \quad y \to x, \quad w \to -w.
$$
Given a surface $\mathcal L$, recall that its {\it evolutes} (also known
as focal surfaces) are the loci of the principal centres of curvature of
$\mathcal L$.
Obviously, a generic surface $\mathcal L$ has two evolutes.
They interchange positions under the change of the orientation.

\begin{proposition}[Ribaucour~\cite{Ri}]
\label{prop:Ri}
Evolutes of surfaces of constant astigmatism are pseudospherical
surfaces.
\end{proposition}

\begin{proof}
Let $\mathbf r(x,y)$ be a surface parametrized by curvature lines.
We use the orthonormal frame $(\mathbf e_1,\mathbf e_2,\mathbf n)$,
where
$$
\mathbf e_1 = \mathbf r_x/u, \quad
\mathbf e_2 = \mathbf r_y/v, \quad
\mathbf n = \mathbf e_1 \times \mathbf e_2.
$$
Then the two evolutes $\mathcal L'$ and $\mathcal L''$ are given by
$$
\mathbf r' = \mathbf r + \frac{\mathbf n}p, \qquad
\mathbf r'' = \mathbf r + \frac{\mathbf n}q,
$$
respectively.
An easy calculation using the Gauss--Weingarten formulas~\eqref{G-W:so3}
shows that
$$
\mathbf r'_x = -\frac{p_x}{p^2} \mathbf n, \qquad
\mathbf r'_y = -\frac{p_y}{p^2} \mathbf n + (1 - \frac qp) \mathbf r_y,
\\
\mathbf r''_x = -\frac{q_x}{q^2} \mathbf n + (1 - \frac pq) \mathbf r_x,
\qquad
\mathbf r''_y = -\frac{q_y}{q^2} \mathbf n,
$$
the unit normals being
$$
\mathbf n' = \frac{\mathbf r_x}u, \qquad
\mathbf n'' = \frac{\mathbf r_y}v.
$$

Now assume $\mathbf r(x,y)$ to be a surface of constant astigmatism.
By applying the substitutions~\eqref{pquv2w} we obtain the first
fundamental form of the evolutes in terms of~$w$:
$$
`I'
   = (w_x\,dx + w_y\,dy)^2 + `e^{-2 w}\,dy^2
   = dw^2 + `e^{-2 w}\,dy^2,
\\
`I''
   = `e^{2 w}\,dx^2 + (w_x\,dx + w_y\,dy)^2
   = `e^{2 w}\,dx^2 + dw^2.
$$
These are the well known pseudospherical metrics in terms of geodesic
coordinates $w,y$ and $w,x$ on the first and the second sheet,
respectively.
\end{proof}

For further reference we also compute the second fundamental forms
$$
`II'
   = -`e^w w_x\,dx^2 + \frac{w_x}{`e^{3w}}\,dy^2,
\qquad
`II''
   = `e^{3w} w_y\,dx^2 - \frac{w_y}{`e^w}\,dy^2.
$$
Proposition~\ref{prop:Ri} provides as with a couple of transformations
from the constant astigmatism equation~\eqref{z} to the sine-Gordon 
equation.
To write them explicitly, we need to equip $\mathcal L'$ and
$\mathcal L''$ with the asymptotic coordinates $\ax,\ay$, i.e.,
the fundamental forms have to be
$$
`I' = d\ax^2 + 2\cos \phi' \,d\ax\,d\ay + d\ay^2,
\quad
`II' = 2\sin \phi' \,d\ax\,d\ay,
\\
`I'' = d\ax^2 + 2\cos \phi'' \,d\ax\,d\ay + d\ay^2,
\quad
`II'' = 2\sin \phi'' \,d\ax\,d\ay.
$$
Here $\phi'$ and $\phi''$ are the angles between the coordinate lines
on $\mathcal L'$ and $\mathcal L''$, respectively.
Using the previous expression of fundamental forms $`I',`II'$ and
$`I'',`II''$ in terms of the variable $w$, we easily see that $\ax,\ay$
can be obtained by the ``reciprocal transformation''~\cite{R-S}
$$
\numbered\label{xieta}
d\ax = \frac12 \sqrt{(w_x + `e^{2 w} w_y)^2 + `e^{2 w}}\,dx
   + \frac12 \sqrt{(`e^{-2 w} w_x + w_y)^2 + `e^{-2 w}}\,dy,
\\
d\ay = \frac12 \sqrt{(w_x - `e^{2 w} w_y)^2 + `e^{2 w}}\,dx
   - \frac12 \sqrt{(`e^{-2 w} w_x - w_y)^2 + `e^{-2 w}}\,dy.
$$
These formulas are valid on both sheets and reflect another property
established by Ribaucour~\cite{Ri}, namely that asymptotic lines on
$\mathcal L'$ and $\mathcal L''$ correspond.

Then the angle $\phi'$ associated with the first sheet satisfies
$$
\numbered\label{phi'}
\cos\phi' = \frac{w_x^2 - `e^{2w} - `e^{4w} w_y^2}
   {\sqrt{(w_x + `e^{2 w} w_y)^2 + `e^{2 w}}
    \sqrt{(w_x - `e^{2 w} w_y)^2 + `e^{2 w}}},
\\
\sin\phi' = -\frac{2 `e^{w} w_x}
   {\sqrt{(w_x + `e^{2 w} w_y)^2 + `e^{2 w}}
    \sqrt{(w_x - `e^{2 w} w_y)^2 + `e^{2 w}}},
$$
while the angle $\phi''$ associated with the second sheet satisfies
a similar set of equations related by the substitution~\eqref{w:ds}.

\begin{proposition}
Let $z(x,y)$ be a solution of the constant astigmatism equation~\eqref{z},
let $w = \frac12 \ln z$.
Determine function $\phi'$ by formula~\eqref{phi'}, and new
coordinates $\ax,\ay$ by the reciprocal transformation~\eqref{xieta}.
Then $\phi'(\ax,\ay)$ is a solution of the sine-Gordon equation 
$\phi_{\ax\ay} = \sin\phi$. 
\end{proposition}

Another solution of the sine-Gordon equation can be obtained by combination
with the discrete symmetry~\eqref{w:ds}. The other symmetries (translation 
and scaling) do not lead to essentially new solutions.

Now, it is easy to check that {\it the evolutes of surfaces of constant
astigmatism are related by the classical Bianchi transformation.}
Indeed, the corresponding points $\mathbf r'$ and $\mathbf r''$ have a
constant distance equal to $1/p - 1/q$.
The corresponding normals $\mathbf n' = \mathbf r_x/u$ and
$\mathbf n'' = \mathbf r_y/v$ are orthogonal.
Finally, being directed along the normal vector $\mathbf n$, the line
joining the points $\mathbf r'$ and $\mathbf r''$ is tangent to both
surfaces $\mathcal L'$ and~$\mathcal L''$.
These three properties characterize the classical Bianchi
transformation.
The Bianchi transformation is, however, superseded by the classical
B\"acklund transformation~\cite{BT}, where the condition on the angle
between the normals is relaxed from being right to being constant.

\section{Surfaces of constant astigmatism as involutes}
\label{involut}

In principle, all surfaces of constant astigmatism can be obtained
from solutions of the sine-Gordon equation as involute surfaces, see, 
e.g., Darboux~\cite[\S802]{Dar}, Bianchi~\cite[\S130--\S150]{BiaI} or
Weatherburn~\cite[Ch.~8]{Wea}.
Geodesic nets on pseudospheric surfaces fall into three classes:
hyperbolic, parabolic, and elliptic~\cite[\S102]{BiaI}.   
Of them only the parabolic geodesic nets lead to surfaces of constant 
astigmatism~\cite[\S136]{BiaI}.   

Recall that the sine-Gordon $\phi_{\ax\ay} = \sin\phi$ describes surfaces 
of constant curvature $-1$ in the asymptotic coordinates~$\ax,\ay$.
By definition,
$$
`I = d\ax^2 + 2 \cos\phi\,d\ax\,d\ay + d\ay^2, \qquad
`II = 2 \sin\phi\,d\ax\,d\ay,
$$
which leads to the Gauss--Weingarten equations
$$
\numbered\label{GW-sG}
\mathbf r_{\ax\ax} = \frac{\cos\phi\,\mathbf r_\ax - \mathbf
r_\ay}{\sin\phi}
   \phi_\ax, \quad
\mathbf r_{\ax\ay} = \sin\phi\,\mathbf n, \quad
\mathbf r_{\ay\ay} = \frac{\cos\phi\,\mathbf r_\ay - \mathbf
r_\ax}{\sin\phi}
   \phi_\ay, \\
\mathbf n_\ax = \frac{\cos\phi\,\mathbf r_\ax - \mathbf
r_\ay}{\sin\phi},
\qquad
\mathbf n_\ay = \frac{\cos\phi\,\mathbf r_\ay - \mathbf
r_\ax}{\sin\phi}.
$$

Recall that coordinates $\gx,\gy$ on a pseudospheric surface are called 
{\it parabolic geodesic} if the first fundamental form can be 
written as
$$
`I = d\gx^2 + `e^{2\gx}\,d\gy^2.
$$
To find the transformation from asymptotic to parabolic geodesic 
coordinates, observe that
$d\ax^2 + 2 \cos\phi\,d\ax\,d\ay + d\ay^2 = d\gx^2 + `e^{2\gx}\,d\gy^2$
is equivalent to the system
$$
\gx_\ax^2 + `e^{2\gx}\gy_\ax =  1, \quad
\gx_\ax \gx_\ay + `e^{2\gx}\gy_\ax \gy_\ay = \cos\phi, \quad
\gx_\ay^2 + `e^{2\gx}\gy_\ay =  1.
$$
This system can be rewritten as
$$
\numbered\label{g2a}
\gx_\ax = \cos\alpha, \qquad\qquad \gy_\ax = `e^{-\gx} \sin\alpha, \\
\gx_\ay = \cos\beta, \qquad\qquad \gy_\ay = `e^{-\gx} \sin\beta,
$$
and 
$$
\numbered\label{pab}
\phi = \alpha - \beta.
$$
In fact,~\eqref{pab} could be also $\phi = \beta - \alpha$, which can 
be reversed by changing the orientation of the surface.
The new unknowns $\alpha$ and $\beta$ can be identified with the angles 
between the geodesics and the two asymptotic coordinate lines.

The integrability conditions of system~\eqref{g2a} are
$$
\numbered\label{ab}
\beta_\ax = -\sin \alpha, \qquad
\alpha_\ay = -\sin\beta,
$$
or, in view of relation~\eqref{pab}, 
$$
\numbered\label{b2a}
\beta_\ax = -\sin(\phi + \beta), \qquad
\beta_\ay = -\phi_\ay - \sin\beta.
$$
These are already compatible by virtue of the sine-Gordon equation.
From equations~\eqref{g2a} we obtain
$$
\mathbf r_\gx = -\frac{\sin\beta}{\sin\phi} \mathbf r_\ax
 + \frac{\sin\alpha}{\sin\phi} \mathbf r_\ay, \qquad
\mathbf r_\gy = (\frac{\cos\beta}{\sin\phi} \mathbf r_\ax
 + \frac{\cos\alpha}{\sin\phi} \mathbf r_\ay)`e^\gx.
$$

With respect to a given geodesic net, the involute surface 
$\tilde{\mathbf r}$ is composed of individual involute curves to the 
geodesics, based on one and the same orthogonal line $\gy = `const$.
Hence, 
$$
\tilde{\mathbf r} = \mathbf r + (a - \gx) \mathbf r_\gx,
$$
where $a$ is an arbitrary constant. 
With the help of equations~\eqref{GW-sG}, the fundamental forms 
$\tilde{`I},\tilde{`II}$ of the involute surface $\tilde{\mathbf r}$ 
can be routinely computed in asymptotic coordinates. 
In particular, the unit normal is $\tilde{\mathbf n} = \mathbf r_\gx$
and
$$
\tilde{`I} = (\gx^2 - \gx + \frac12) (1 - \cos 2\alpha)\,d\ax^2
 + (2\gx - 1) (\cos(\alpha + \beta) - \cos\phi)\,d\ax\,d\ay 
\\\quad
 + (\gx^2 - \gx + \frac12) (1 - \cos 2\beta)\,d\ay^2, 
\\
\tilde{`II} = (\gx - \frac12) (\cos 2\alpha - 1)\,d\ax^2
 + (\cos(\alpha + \beta) - \cos\phi)\,d\ax\,d\ay 
\\\quad
 + (\gx - \frac12) (\cos 2\beta - 1)\,d\ay^2.
$$
Hence, the principal radii of curvature are $\gx$, $\gx - 1$.
The Gauss--Mainardi--Codazzi equations of the involute surface hold 
as a consequence of the sine--Gordon equation, the two 
equations~\eqref{g2a} on $\gx$ and the system~\eqref{b2a} on~$\beta$.  

To obtain the corresponding solution of the constant astigmatism 
equation~\eqref{z}, we have to reparametrize the involute surfaces by 
curvature lines.
Let $x,y$ denote the new coordinates. We choose $x = \gy$ and 
define $y$ by the compatible system of equations
$$
\numbered\label{y2a}
y_\ax = `e^\gx \sin\alpha, \qquad
y_\ay = `e^\gx \sin\beta.
$$
A routine calculation shows that $`e^{-2\gx(x,y)}$ is a solution of 
the constant astigmatism equation~\eqref{z}. Summarizing, we have the 
following proposition.

\begin{proposition}
Let $\phi(\ax,\ay)$ be a solution of the sine-Gordon equation
$\phi_{\ax\ay} = \sin\phi$.
Let $\alpha,\beta$ be solutions of the compatible equations
$$
\beta_\ax = -\sin \alpha, \qquad
\alpha_\ay = -\sin\beta, \qquad
\alpha - \beta = \phi.
$$
Determine functions $\gx,x,y$ by equations
$$
d\gx = \cos\alpha\,d\ax + \cos\beta\,d\ay, \\
dx = `e^{-\gx} (\sin\alpha\,d\ax + \sin\beta\,d\ay), \\
dy = `e^\gx (\sin\alpha\,d\ax + \sin\beta\,d\ay).
$$
Then the function $`e^{-2\gx(x,y)}$ is a solution of the constant 
astigmatism equation~\eqref{z}.
\end{proposition}

\begin{example} \rm
{\it Von Lilienthal's surfaces} (involutes of the pseudosphere).
Published in 1887, these surfaces seem to have fallen into oblivion.
Recall that the pseudosphere is a surface obtained by rotating the
tractrix around its asymptote. The meridians are geodesics of the 
parabolic type and therefore von Lilienthal's surface is obtained by 
rotating the involute of the tractrix (which itself is the involute 
of the catenary).

In geodesic coordinates $\gx,\gy$, the ``upper half'' of the pseudosphere
has a parametrization
$$
\mathbf r = (\begin{array}{c}
  `e^{-\gx} \cos \gy \\ `e^{-\gx} \sin \gy \\
  `arcosh`e^{\gx} - \sqrt{1 - `e^{-2 \gx}}
  \end{array}), \quad \gx > 0,
$$
whose first fundamental form is $d\gx^2 + `e^{-2 \gx}\,d\gy^2$ (differs by 
the sign of $\gx$ from the canonical form used in the preceding section).
Then
$$
\tilde{\mathbf r} = \mathbf r + (a - \gx) \mathbf r_\gx
  = (\begin{array}{c}
    (\gx - a + 1) `e^{-\gx} \cos \gy \\
    (\gx - a + 1) `e^{-\gx} \sin \gy \\
    `arcosh`e^\gx - (\gx - a + 1) \sqrt{1 - `e^{-2 \gx}}
\end{array}), \quad \gx > 0,
$$
parametrizes a rotational surface, for every real constant $a$.
The surface is regular for all $a \le 0$.
Otherwise it has a cuspidal edge at $\gx = a$, which is a circle of radius
$`e^{-a}$.
Another singularity that occurs for every $a > 1$ is the intersection
with the rotation axis at $\gx = a - 1$.
Choosing the orientation so that the normal vector is
$$
\tilde{\mathbf n} = (\begin{array}{c}
    -`e^{-\gx} \cos \gy \\
    -`e^{-\gx} \sin \gy \\
    \sqrt{1 - `e^{-2 \gx}}
\end{array})
$$
(i.e., $\mathbf n$ swaps orientation when crossing either of the
singularities),
then
$$
\tilde{`I} = \frac{(\gx - a)^2}{`e^{2 \gx} - 1}\,d\gx^2
  + \frac{(\gx - a + 1)^2}{`e^{2 \gx}}\,d\gy^2,
\\
\tilde{`II} = \frac{\gx - a}{`e^{2 \gx} - 1}\,d\gx^2
  + \frac{\gx - a + 1}{`e^{2 \gx}}\,d\gy^2.
$$
and the principal radii of curvature are $\gx - a$ and $\gx - a + 1$.
The corresponding solution of the constant astigmatism equation~\eqref{z} 
is
$$
z = \frac1{x^2 - `e^{2(a - 1)}}.
$$

\setlength{\unitlength}{1.4in}
\newcommand{\obr}[3] 
{
\begin{picture}(1,1)(0,0)
\put(0.6,0.5){\makebox(0,0){\includegraphics[width=3in]{#2}}} 
\put(0.6,0.5){\makebox(0,0){\vrule depth 0cm height 1.6in width .05mm}} 
\put(0.6,0.5){\makebox(0,0){\includegraphics[width=3in]{#1}}} 
\put(0.6,0){\makebox(0,-0.35){\small $a = #3$}} 
\end{picture}
}

\begin{figure}[h]
\vskip 1pc
\begin{center}
\obr{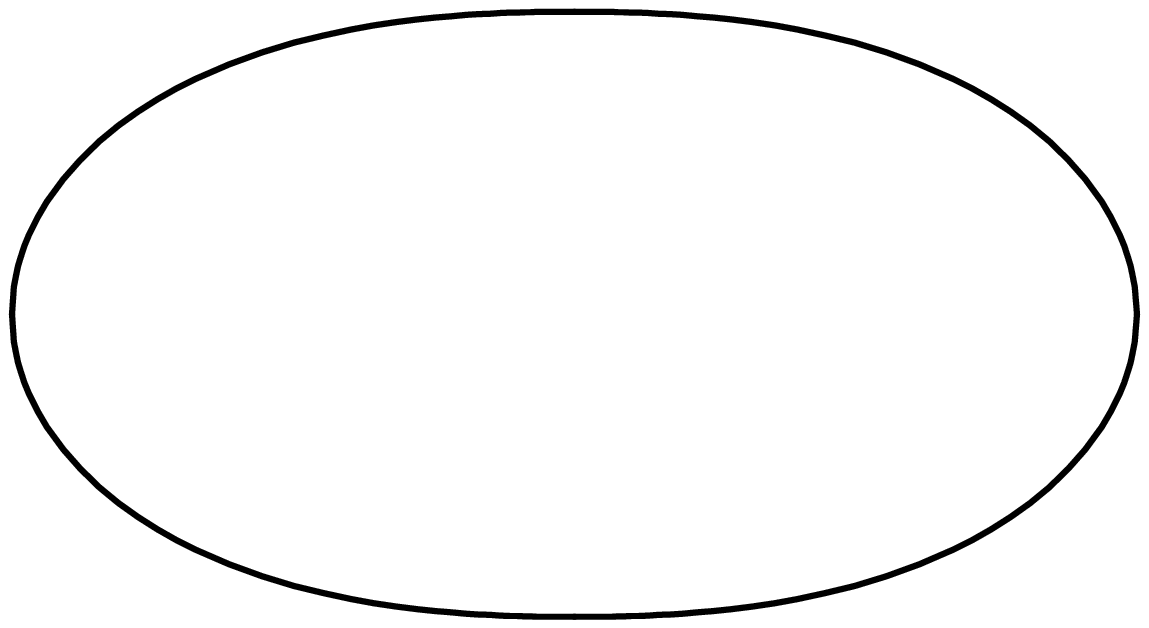}{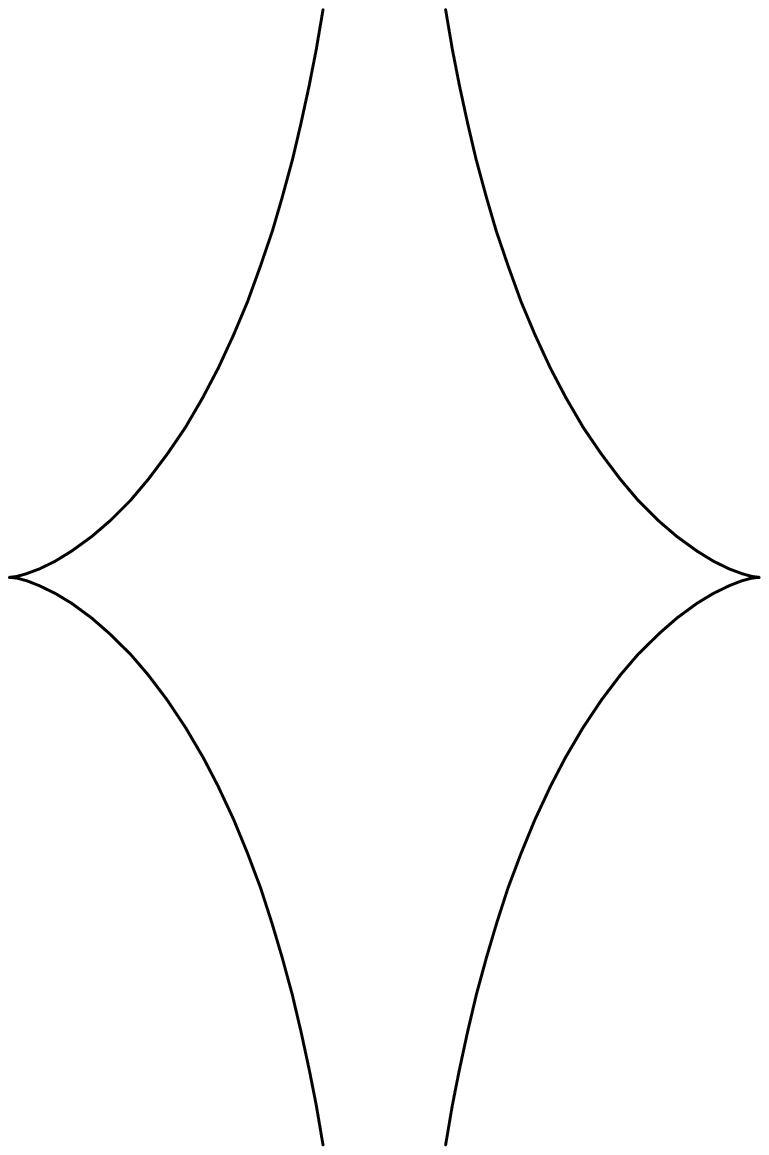}{-0.5}
\obr{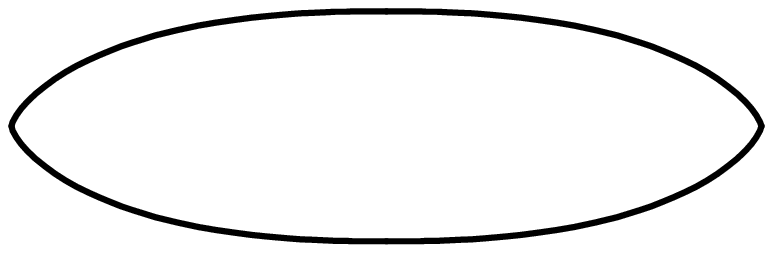}{plot_common.ps}{0}
\obr{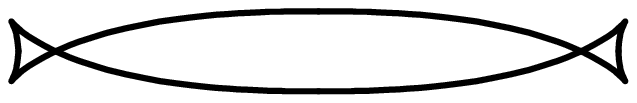}{plot_common.ps}{0.2}
\obr{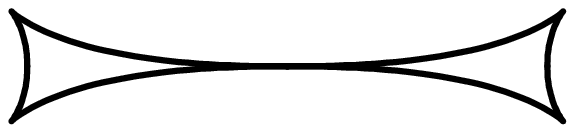}{plot_common.ps}{1-\ln 2}

\vspace{0.7in}
\obr{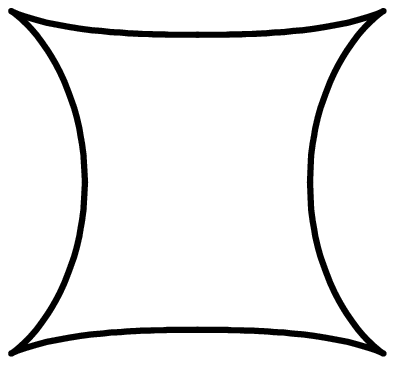}{plot_common.ps}{0.7}
\obr{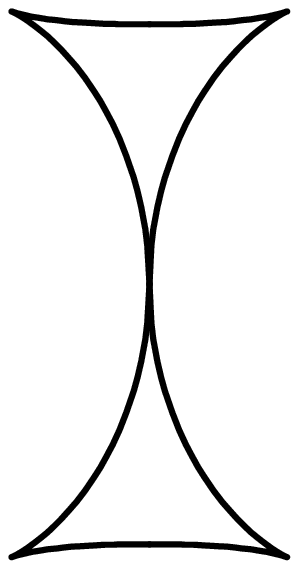}{plot_common.ps}{1}
\obr{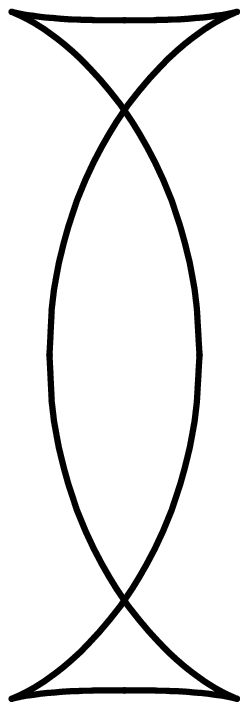}{plot_common.ps}{1.2}
\obr{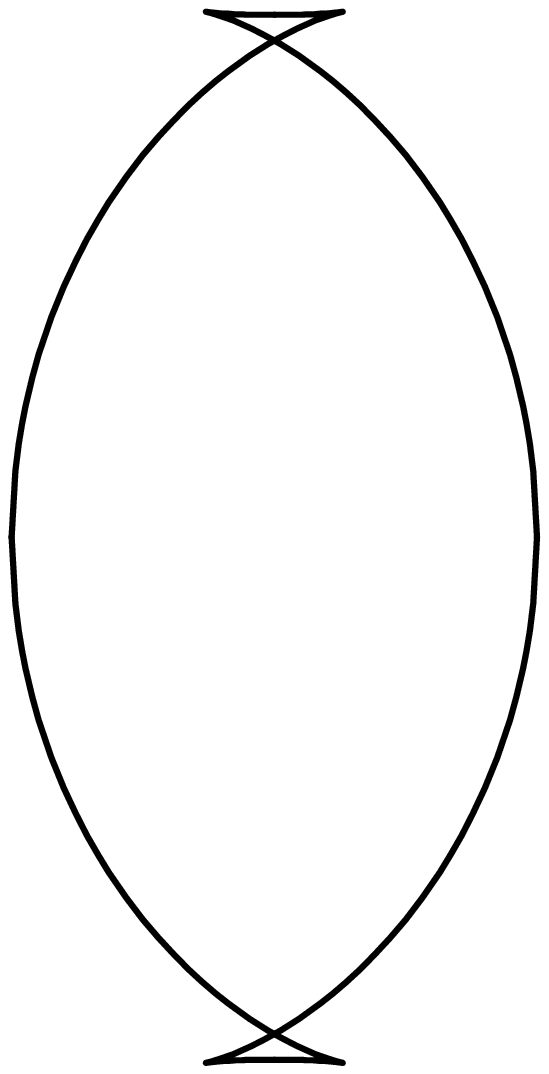}{plot_common.ps}{1.7}
\end{center}
\vskip 1pc
\caption{A gallery of von Lilienthal surfaces} \label{Pictures}
\end{figure}

Plane sections of von Lilienthal surfaces for various values of the parameter 
$a$ can be seen on Figure~\ref{Pictures}. 
Besides the rotation axis, each picture shows the tractrix, which is the 
plane section of the pseudosphere, and its involute curve, which is the plane 
section of the von Lilienthal surface. 

We finish this example with a short exploration of the behaviour at the 
limits of the definition domain.
For $\gx = \infty$ the surface closes up at a point on the rotation axis 
at the height $a - 1 + \ln 2$, where both principal radii of curvature 
are infinite (the zero height is that of the cusp of the tractrix).
For $\gx \to 0$ the surface vertically approaches a horizontal circle of 
diameter $|1 - a|$.
Two surfaces $\tilde{\mathbf r}(\gx,\gy)$ and $-\tilde{\mathbf r}(\gx,\gy)$
can be glued along this circle to form a single surface of constant
astigmatism~$1$.
For $a = 1$ both glued surfaces have a cusp here.
\end{example}

\section{Conclusions and discussion}

Among the still incomplete results of classification of integrable
Weingarten surfaces, we have identified a class originally introduced
and
investigated by nineteenth-century geometers. 
The class, which we propose to call surfaces of constant astigmatism,
is governed by the equation
$$
z_{yy} + (\frac1z)_{xx} + 2 = 0.
$$
For this equation we found an $\mathfrak{sl}(2)$-valued zero curvature
representation depending on a parameter, a third-order symmetry, and a
nonlocal transformation to the sine-Gordon equation
$\phi_{\ax\ay} = \sin\phi$.
We had to leave aside the problem of finding a B\"acklund transformation
as well as a recursion operator producing a hierarchy of local
symmetries. 

It should be stressed that the classification problem of integrable 
surfaces is far from being easy. An obvious reason lies in the abundance 
of integrability-preserving ways to derive one surface from another. 
Clearly, parallel surfaces, evolutes, and involutes of integrable 
surfaces are integrable. On the differential equation level, the 
corresponding notion is that of the covering~\cite{K-V}.
The integrable classes of surfaces must be either closed with respect to
taking derived surfaces or the derivation must map one integrable class 
into another.

\ack

This paper would be impossible without encouragement, support and advice
from J.~Cie\'sli\'nski, E. Ferapontov, R. L\'opez and A. Sergyeyev.
The first-named author was supported by GA\v{C}R under project
201/07/P224.
The second-named author by M\v{S}MT under project MSM~4781305904.
Thanks are also due to CESNET for granting access to the MetaCentrum
computing facilities.

\end{document}